# Successful High Power Acceleration of the HSC Type Injector for Cancer Therapy in IMP

LU Liang [1], HATTORI Toshiyuki [2], ZHAO Huan-yu [1], KAWASAKI Katsunori [3], SUN Liepeng [1], JIN Qianyu [1], ZHANG Jun-jie [1], SUN Liangting [1], HE Yuan [1], ZHAO Hong-wei [1]

[1] Institute of Modern Physics, Chinese Academy of Sciences, Lanzhou 730000, China.

[2] National Institute of Radiological Sciences, Chiba 263-8555, Japan.

[3] Tokyo Institute of Technology, Tokyo 152-8550, Japan

Abstract: A Hybrid single cavity (HSC) linac, which is formed by combining a radio frequency quadrupole (RFQ) structure and a drift tube (DT) structure into one interdigital-H (IH) cavity, is fabricated and assembled as a proof of principle (PoP) type injector for cancer therapy synchrotron according to the culmination of several years' researches [1-4]. The injection method of the HSC linac adopt a direct plasma injection scheme (DPIS), which is considered to be the only method for accelerating an intense current heavy ion beam produced by a laser ion source (LIS). The input beam current of the HSC was designed to be 20 milliampere (mA) $C^{6+}$ ions. According to numerical simulations, the HSC linac could accelerate a 6-mA $C^{6+}$ beam which meets requirement of the needed particle number for cancer therapy use ($10^{8\sim9}$ ions/pulse). The injection system adopted HSC injector with the method of DPIS can make the existing multi-turn injection system and the stripping system unnecessary, and also can make the beam pipe of the existing synchrotron magnets downsize which could reduce the whole cost of synchrotron (estimated millions dollars). Details of the measurements and evaluations of the assembled HSC linac, and the results of the high power test using a LIS with DPIS are reported in this paper.

[1] Corresponding author's address: Linear accelerator group, Accelerator Div., IMP, 509 Nanchang Rd., Lanzhou, Gansu730000, China.

Tel: +86-391-4969614. E-mail: luliang@impcas.ac.cn (Liang LU).

Work supported by the National Natural Science Foundation of China (Grant No. 90Y436070) and One Hundred Person Project of CAS (Grant No. 1103Y536060)

# 1 Introduction

Recently, heavy ion cancer therapy has proved to be a remarkably effective treatment. However, the injection accelerators of existing facilities are large in size and high in cost. The first generation of heavy ion cancer therapy facility, i.e., heavy ion medical accelerator (HIMAC) in Chiba that has cured over five thousand patients [5], is over 30 meters in length for the linear accelerator part [6]. And the length of linear injector of a new generation of heavy ion cancer therapy facility at Gunma University is over 6 meters, which began treatments in 2010 [7]. The electron cyclotron resonance (ECR) source used for existing heavy ion synchrotron facilities could produce only hundreds of microampere $C^{4+}$ ions, thus the stripping system, which could make the beam emittance growth bigger, and the multi-turn injection system are necessary for changing ion state form $C^{4+}$ to $C^{6+}$ and injecting enough ions to the synchrotron. And the accelerating system must accelerate $C^{4+}$ ions up to 6-8 MeV/u for obtaining over 90% charge-state changing ratio. In our research, we aimed to design a compact injection linac that was less than two meters in length but had an ability to directly accelerate the high intensity $C^{6+}$ ion beams by using a LIS with DPIS. Our new injector, named the HSC linac, which can accelerate $C^{6+}$ ions up to 2.1 MeV/u (25 MeV for $C^{6+}$ ions) from 25 keV/u in 1800 mm with less than 100 kilowatts (kW) of power, has been designed and fabricated shown in the Fig. 1. The details of the designs and simulations could be found in the series papers [1-4].

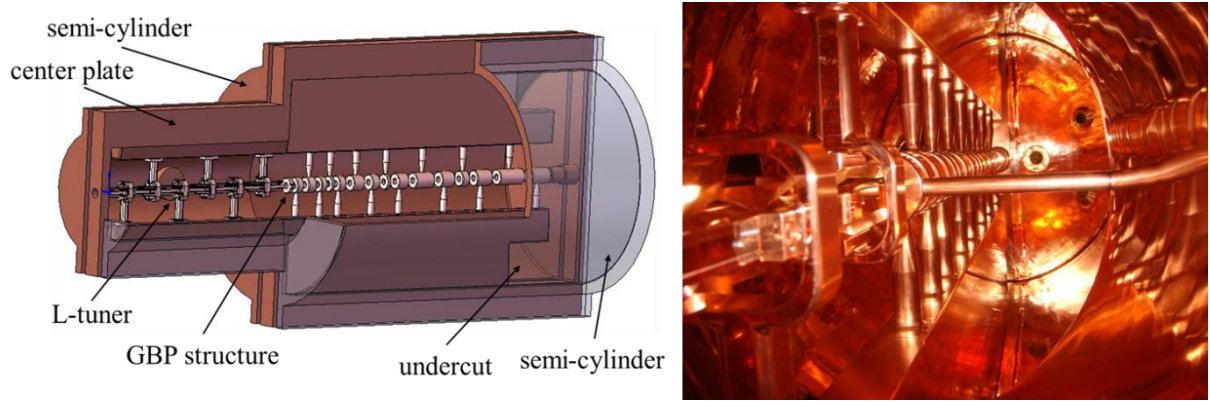

Fig. 1: Before and after assembly images of the HSC linac. The HSC linac was assembled by using a sandwich method (a center plate, two semi-cylinders).

The shunt impedance calculated based on perturbation measurements was 115.4 M/m, which is 94.4% of the MWS simulated value of 122.2 M/m. The 94.4% of the shunt impedance percentage is very close to 95% of the Q value percentage. The shunt impedance value of 115.4 M/m for the HSC structure cavity is a quite higher value when compared to other linac structures within the same beam velocity region, as shown in the Fig. 2 [8-9]. According to simulations, the HSC linac could accelerate a 5.98 mA $C^{6+}$ ion beam, which contains sufficient ion numbers for cancer therapy. Thus, it can be used directly as an injector without the stripper system and the multi-turn injection system. The results of the pretest and the low power test, including the resonant frequency, electric (E) field distribution of the cavity, and the tuning effect of the L-tuner match well with the numerical calculations. The measured frequency of the cavity was 100.49 MHz, which was within 0.5% of the designed value of 100 MHz.; the measured best Q value was 95% of the simulated value, which was considered as a new world record. And last year, the HSC linac was transported to the Institute of Modern Physics (IMP) [10] of Chinese Academy of Sciences from Japan. The IMP could provide a 250 kW radio frequency (RF) power source and a high current LIS [11] for further HSC research. In the last few months, the high power $C^{6+}$ beam accelerations were operated and analyzed. The results of the high power test matched the simulations and calculations well. That implies our research achieved successes.

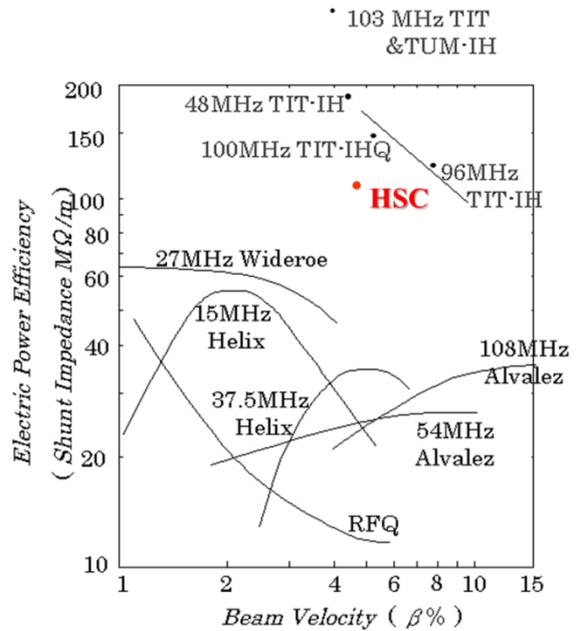

Fig. 2: A comparison image of the shunt impedance of the HSC structured linac compared with that of other linac structures.

## 2 DPIS Test and High Power Test in IMP

### 2.1 Injection System and DPIS Test

The injection system is comprised of a laser generation machine, a chamber for plasma generation, a solid carbon target, an optical system and a vacuum system. The laser generation is a Nd; YAG laser (Beamtech Optronics Co., Ltd., S.#: SGR-30) with a Q switch to generate the laser ablation plasma, which could provide a maximum 20 mA C6+ ion beam with a maximum 60 kilo-voltage (kV) extraction voltage. The wave length of the generated laser is 1064 nm. The maximum laser power is 3 joule (J) from the laser machine, the laser energy and the power density at the focal point on the target is 1.9 J and $6.7 \times 10^{12}\ W/cm^2$. The target is a plate type solid target (TOYO

TANSO (Japan) Co., Ltd., purity: 99.999%, height: 100 mm; width: 50 mm; thickness: 5 mm). A voltage of 50 kV (Spellman high voltage electronics Co., Ltd., S.#: SL1200) is applied to the inner high voltage platform which connects with a plasma nozzle in order to inject $C^{6+}$ ions at a designed input energy of 25 keV/u into the grounded HSC linac. The generated plasma passes through the plasma nozzle at 50 kV and comes out from a nozzle tip. There is a length-adjustable slit, installed on the tip, with a 7 mm inner diameter. The distance from target to extraction nozzle is 826 mm, and the distance from nozzle to the RFQ entrance is 12 mm. The illustration of this injection system for the HSC lianc is shown in the Fig. 3. This system could inject 17.5 mA $C^{6+}$ ions to RFQ rods. The charge states of produced ions are shown in the Fig .4

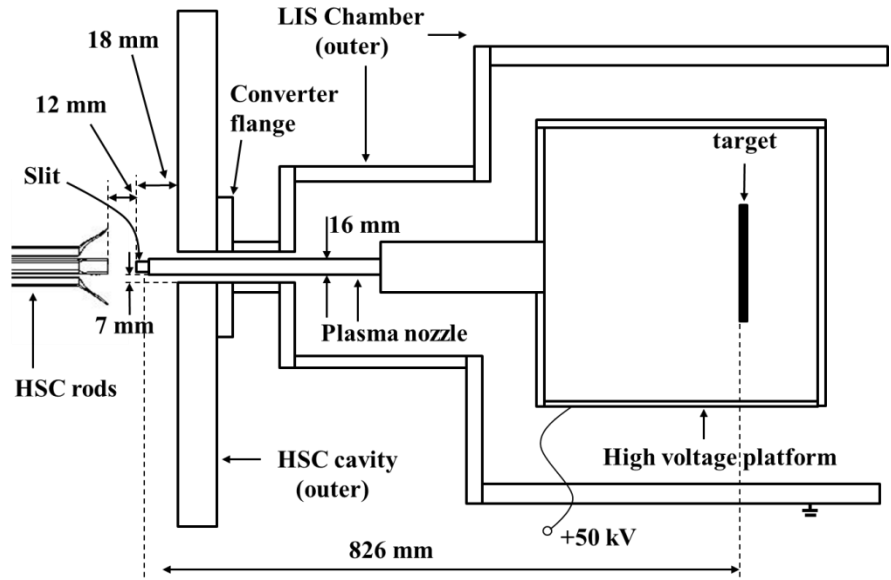

Fig. 3: An image of the injection system for the HSC linac.

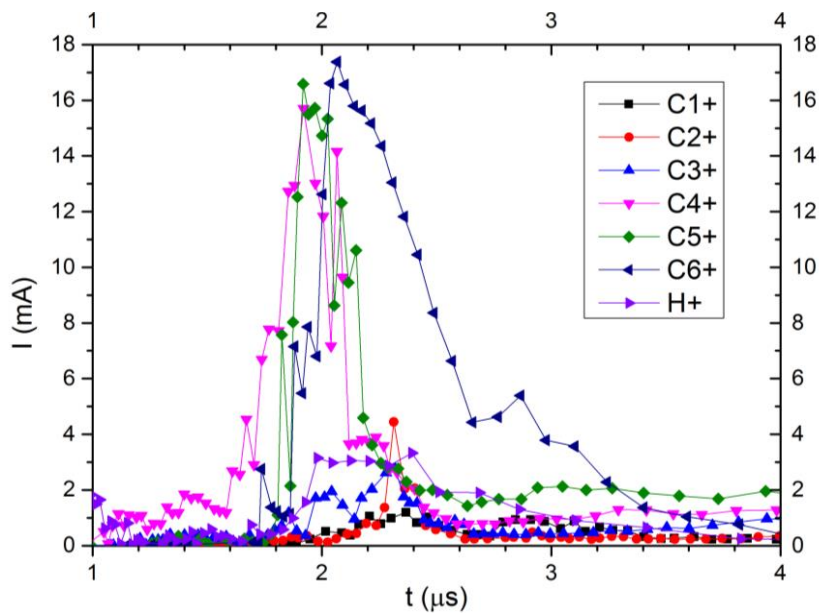

Fig. 4: An image of the charge states from LIS, calculated position at the extraction slit shown in Fig. 3.

## 2.2. High power RF tests

Two type coaxial waveguide (WX-152D in the RF source side and WX-77D in the power coupler side), with a 50 Ω impedance, were used for the transmission of the RF power. Directional couplers were used for monitoring the incident and reflection wave. Two pickup antennas, one is located in the DT section and one is located in the RFQ section, were used to measure the input RF power. We aged the cavity while monitoring the temperature of the cooling water and the pressure in the cavity.

The aging of the cavity was performed with a duty factor of 1% until the incident power reached about 25 kW, 0.5% until the incident power reached about 80 kW, 0.2% until the incident power reached about 105 kW, 0.1% until the incident power reached about 110 kW. Finally the incident RF power reached 165 kW and the resonated frequency changed -27 kHz when comparing the initial state of no RF power.

## 3. Beam Acceleration Tests.

### 3.1. Test system

Shown in the Fig. 5, the trigger A, the trigger B and the trigger CD from signal generator 1 (SG1) were used to trigger the laser, Q switch and the RF power, respectively. The trigger B was delayed 680 microsecond (μs) to the trigger A and 400 μs to the trigger CD, and the time length of the trigger CD is 1 millisecond (ms). In our operation, the trigger B was considered as the start point, and the trigger B was almost same to the diode (ALPHALAS GmbH, S.#: 33034) trigger which was used to check the appearance of the laser. The conditions of the RF operation were 1 ms in the pulse width and 0.1% in the duty factor. The target was manually changed the position after each shot. A beam acceleration system, the layout of the system shown in the Fig. 6 was built to measure the beam current. The systems used three turbomolecular pumps to keep the vacuum pressure better than $7.5 \times 10^{-5}$ Pascal (Pa). The exhaust velocity are 1300 L/s (KYKY technogogy Co., Ltd., S.#: FF-200/1300) at the plasma chamber side, 750 L/s (Agilent, S.#: TV750PUMP) at the HSC cavity side, 300 L/s (Agilent, S.#: TV301NAV) at the detector side.

A faraday cup (FC), 40 mm in diameter, a located at 2936 mm downstream from the target surface, was set to measure the total current after acceleration. This FC is an ordinary faraday cup without abilities to measure the micro-bunchers of the accelerated beams and to accurate the peak times to μs level. Two solid state detectors (SSD), one located at the straight beam line after the magnet one located at 45 degree from beam line behind the magnet, were also used to measure the beam signals. As shown in the Fig. 6, a pre-Amp (TENNELEC CO., LTD, model 2248), a bias power source (TENNELEC Co., Ltd., model TC953), a main Amp (TENNELEC CO., LTD, model TC241) and an AD convertor (a multi-channel analyzer, LABO CO., LTD, model 2100C/MCA) were used to transfer the signal from the SSD to a computer. A power crate (W-IE-NE-R, Plein & Baus GmbH, S.#: NIMpact 300) was used to supply for the NIM productions. And two location, one is the exit of the HSC cavity (mesh 1) one is behind the FC (mesh 2), were set meshes to cut off the ions. The transmission of mesh 1 (Wave Corporation, C#20) and mesh 2 (Eggs Corporation, two pieces of #100 mesh and one piece of #60 mesh) are 50% and 5%, respectively. The whole environmental temperature was controlled at 24 Centigrade. The image of the whole high power acceleration system is shown in the Fig. 7.

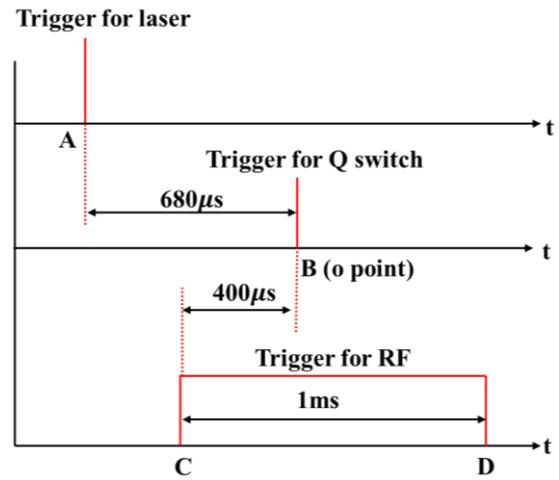

Fig. 5: The triggers from the SG1.

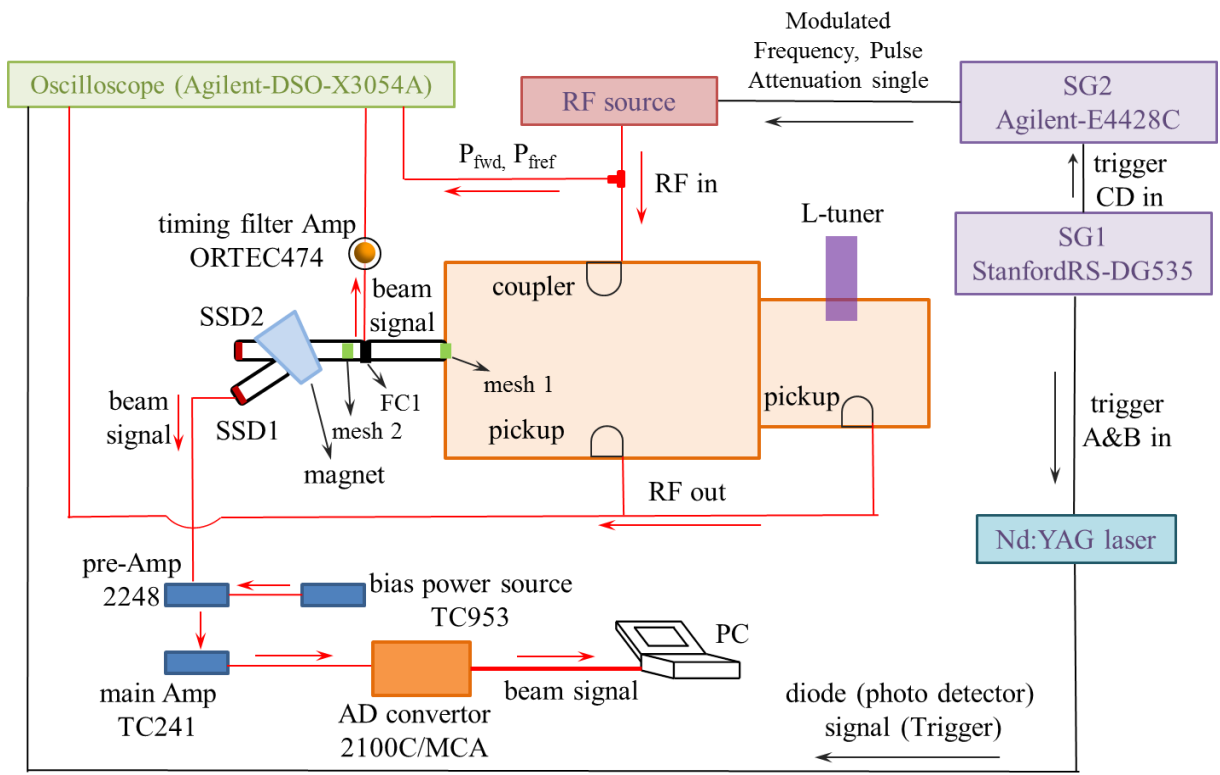

Fig. 6: Layout of the beam acceleration system. The FC, SSD1, SSD2, SG1 and SG2 refer to a faraday cup, solid state detector 1 and solid state detector 2, signal generator 1 and signal generator 2, respectively.

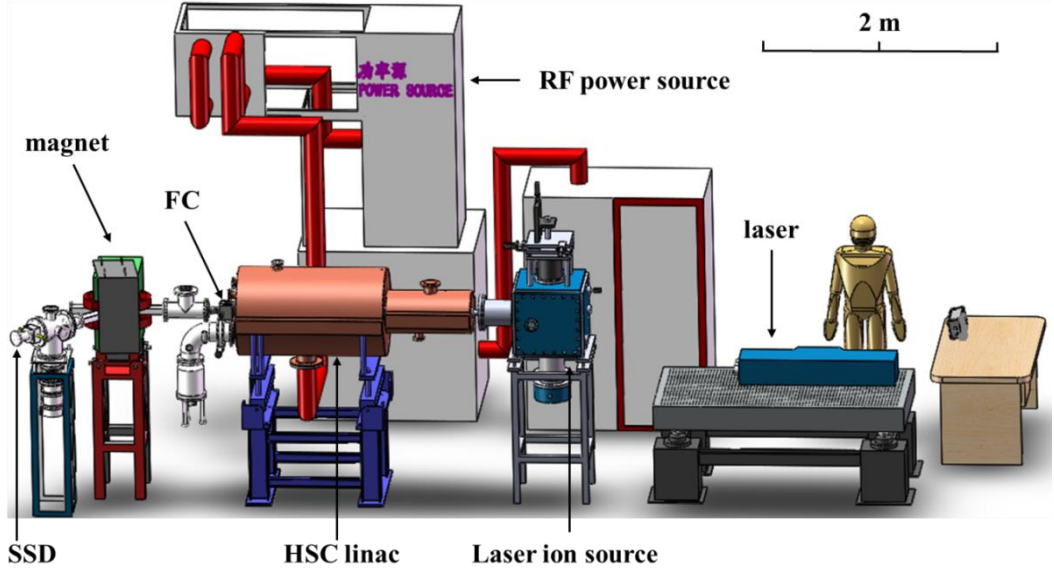

Fig. 7: The image of the high power test system for HSC $C^{6+}$ ion accelerations.

## 3.2 Preliminary experiment

The ability of HSC was designed as $C^{6+}$ (q/A = 1/2) ion accelerations, and the designed Kilpatrick factor was 1.8, thus, the HSC could only accelerate $C^{6+}$ ions and $H^+$ ions. The total accelerated current could be measured by the FC. The injection point, as shown in the Fig. 3, is closer to the rods, the incident and the measured ions number are larger. However, in our injection system, the distance, 12 mm from injection slit to the rods, is a shortest distance. There will be discharge if the distance between the tip of the RFQ electrode and the input surface of the plasma is shorter than 12 mm. In order to match the input energy to the input condition of the RFQ, we measured the proton beam currents with different extraction voltage changing in the ion source. The compatibility condition of the input energy for the HSC linac is 25 keV/u, and a highest accelerated $H^+$ peak is measured at an extraction voltage of 25 kV with a RF power of 27 kW. Shown in the Fig. 8, the peak appears from the extraction voltage of 20 kV, and disappears from the extraction voltage of 30 kV. That implies a voltage of 25 kV is exactly the extraction voltage for $H^+$ beam injection, and the $C^{6+}$ beam commissioning was conducted with the extraction voltage of 50 kV and the RF power of 108 kW. The Fig. 9 shows a shape of the drifted ions without extraction voltage and RF power, a shape of the accelerated $C^{6+}$ ions with 50 kV extraction and 110 kW RF power. The accelerated signal was integrated 100 ns by a timing filter as shown in the Fig. 6. The peak of the accelerated $C^{6+}$ ion reaches 5 mA, which agreed well with simulations and transmission calculations. The numerical transmission could be easily figured out as 5.03 mA by using the formula:

$$T = \left(\frac{L_1}{L_2}\right)^3 \times I \times T_s$$

Here, the $L_1$ implies the existing length form target surface to the injection slit, which would be 826 mm; the $L_2$ implies the length form target surface to the RFQ rods, which would be 838 mm; the I implies the injection current of the $C^{6+}$ ions at the position of the $L_1$, which would be 17.5 mA as shown in the Fig. 4; the $T_s$ implies the simulated transmission, which would be 30% [4].

The calibration of energy channel for SSD was used a radioactive source: $^{241}$Am. The $^{241}$Am could mainly emit alpha particles with an energy of 5.486 MeV (85%) [12]. In our preliminary experiments, the SSDs and the $^{241}$Am were installed and in a vacuum chamber over 30 minutes. The results of the calibration will be presented with the commissioning results below.

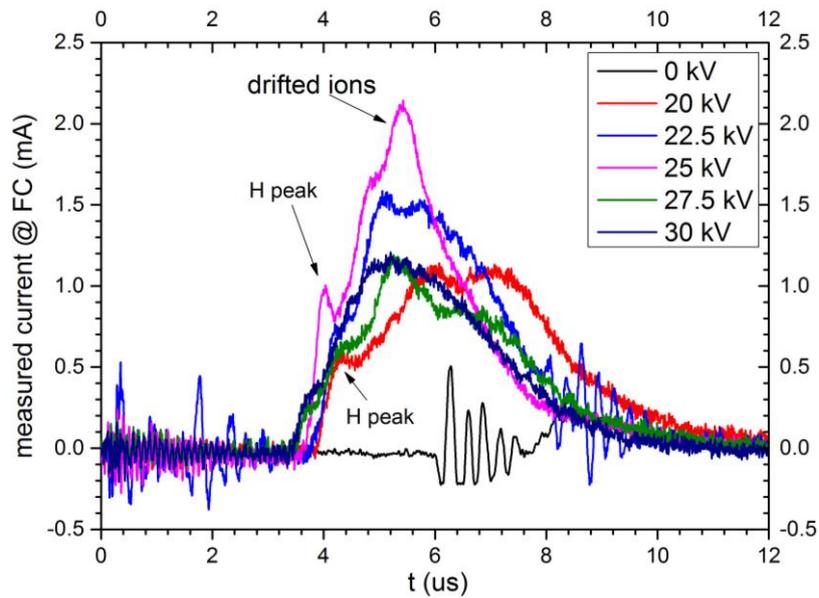

Fig. 8: Measured beam signals with respect to the extraction voltages (plasma input position: 12 mm, RF power: 27 kW).

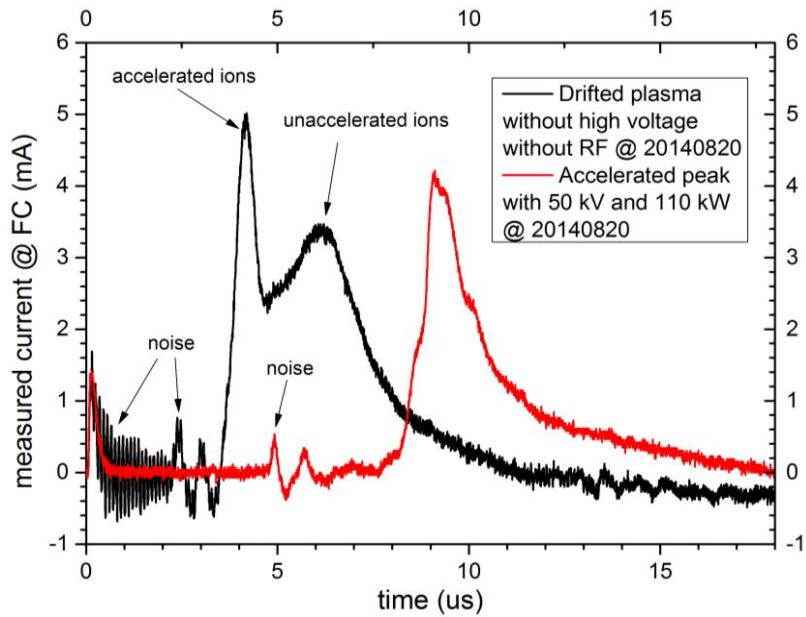

Fig. 9: A measured signal of the accelerated and drifted $C^{6+}$ ion beam current.

## 3.3 Commissioning

As mentioned before, our FC do not have abilities to detect the precise time and the bunch structure of the accelerated ions, we used two SSD type detectors to measure the beam energy. The SSDs (one was made by CANBERRA CO., LTD., model: PD50-12-100AM, S/N: 29167, another SSD was made by OXFORD CO., LTD., model: IPC-50-100-15CB, S/N: 17947-14) were alternately installed on the beam axis downstream of the 45-degree-bending magnet. These two SSDs were installed on the center of a 2-inch flange.

The bending magnet could provide a maximum 1 tesla magnetic field with a maximum 300 A input current. The deflection radius of the magnet is 50 cm. Using SSD, we checked the energy of the accelerated ions firstly by changing the input current of magnet while the operations with 110 kW RF power. As shown in the Fig. 10, there are 4 spectra by bending the accelerated beams. The highest one of them stands on the 8.43 kilogauss where implies the ions with an energy of 25.7 MeV. This energy agreed well with the designed energy of 25 MeV 4.

And as shown in a) and b) of the Fig. 11, both the Canberra and the Oxford SSDs confirmed the accelerated $C^{6+}$ ion beams under operations with the average 110 kW of RF power, and showed the $C^{6+}$ beam energy is 25.7 MeV (2.14 MeV/u). Both the detected results of the beam energy agree very well with each other, and agree very well with the bended results shown in the Fig. 11.

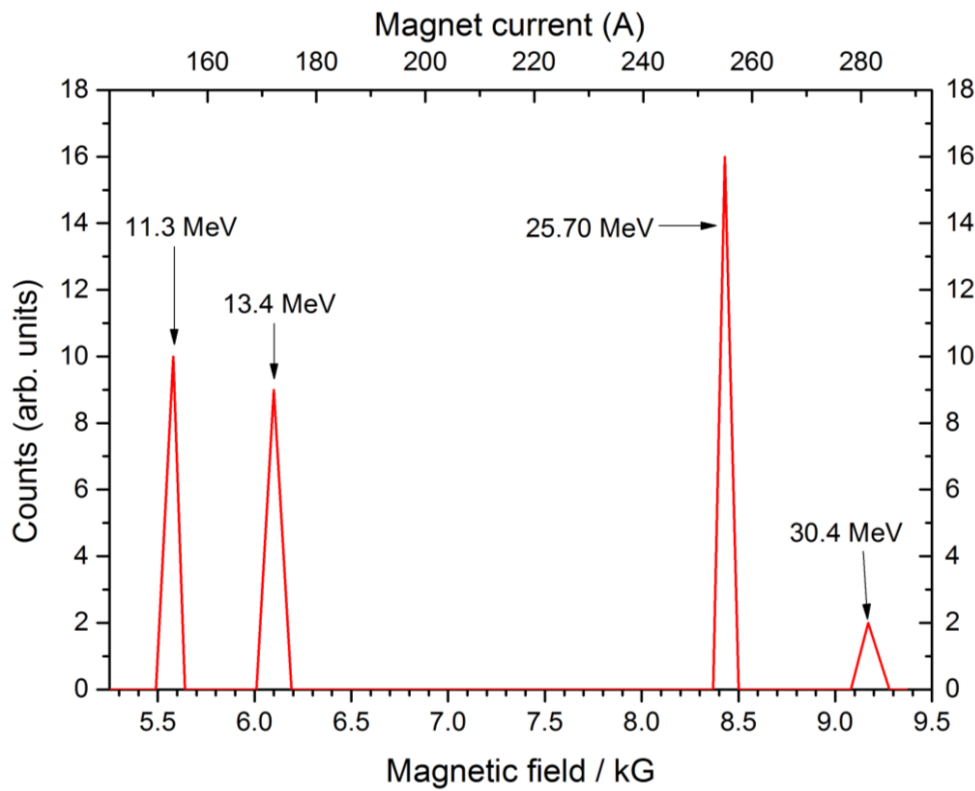

Fig. 10: The spectra show the energy distribution by ions moving through a bending magnet.

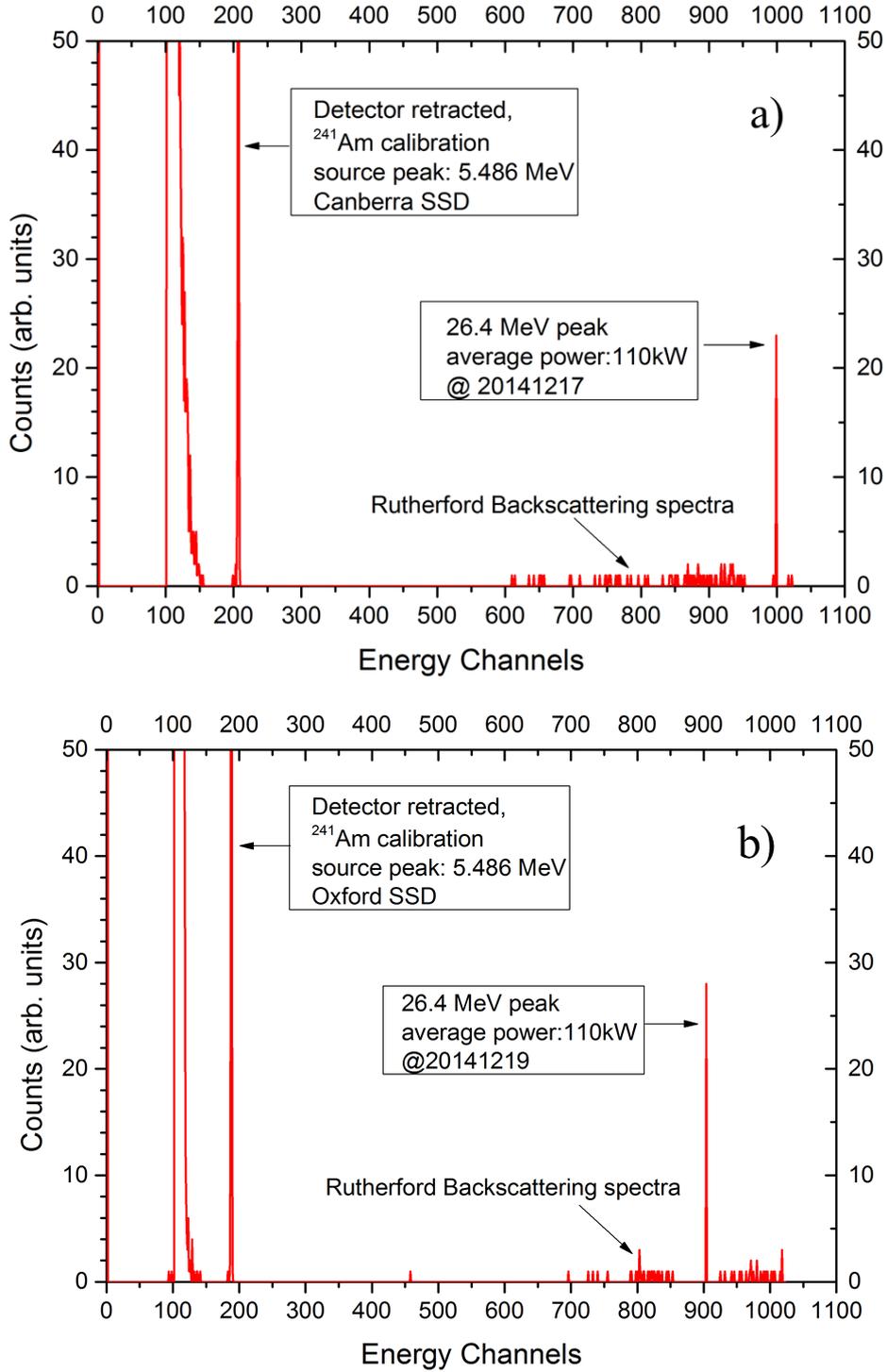

Fig. 11: The Canberra and oxford SSD measured energy channels. The 5.486 MeV channel is calibration source. The detected high signal was calculated as 25.7 MeV.

## 4 Conclusions

The HSC linac as a high intensity heavy ion injector has been proposed to accelerate and inject $C^{6+}$ ions to a synchrotron for cancer therapy. In this study, we manufactured a 2-meter long HSC linac as a prototype hybrid cavity, and successfully commissioned high power tests for $C^{6+}$ ion acceleration. The results of the preliminary tests and the commissioning tests agree well with the designs and the calculations. Further, we also established an engineering process for the design and the manufacture of the HSC cavity.

The successes of the high RF power test proved the HSC have a reliable ability as an injector for heavy ion cancer therapy facilities. Using the HSC, not only the existing stripper system and the existing multi-turn injection system could be cut down, but the magnets of the synchrotron could be reduced to two third or half because the one turn injection could make the size of the existing beam pipe (~200 mm) down to dozens of millimeter. This will save several million dollars for one heavy ion cancer therapy facility, and lighten the burdens for millions patients.

The successes of the high RF power test for the PoP HSC linac also encourage us to develop a new HSC which could use as a real injector for heavy ion cancer therapy facility. We have proposed a series of funds to develop a new machine and one of them was permitted. The HSC linac system using the laser ion source with DPIS would provide new developments for such a heavy ion cancer therapy facility.